\documentclass[12pt]{article}
\usepackage{amsmath,amssymb}
\usepackage{bm}
\usepackage{mathrsfs}
\usepackage{fourier}
\usepackage{multirow}
\allowdisplaybreaks[4]
\newcommand{{\Slashp}}{p\!\!\!\!\!\big/}
\newcommand{{\Slashq}}{q\!\!\!\!\!\big/}
\newcommand{\LBox}{\mbox{\large$\Box$}}
\usepackage[usenames]{color}

\begin{document}

\title{Release of physical modes\\
from unphysical fields}

\author{
Yoshiharu \textsc{Kawamura}\footnote{E-mail: haru@azusa.shinshu-u.ac.jp}\\
{\it Department of Physics, Shinshu University, }\\
{\it Matsumoto 390-8621, Japan}\\
}

\date{
September 1, 2014}

\maketitle
\begin{abstract}
We present a basic idea and a toy model 
that physical modes originate from unobservable fields.
The model is defined on a higher-dimensional space-time
and has fermionic symmetries that make fields unphysical,
and observable modes can appear through a dimensional reduction.
\end{abstract}


\section{Introduction}
\label{Introduction}

The existence of the standard model fields 
leads to the following basic questions, concerning
the structure of the model.
Why is the gauge group $SU(3) \times SU(2) \times U(1)$?
Why exist three families of quarks and leptons?
What is the origin of the weak scale?
It is hard to answer them completely 
without some powerful guiding principles
and/or a more fundamental theory.
What we can do at present is 
to simplify the questions and to find possible solutions
based on some conjectures.

We adopt the fantastic idea 
that our world comes into existence 
from $\lq\lq$nothing'', as a conjecture~\cite{YK1}.
Nothing here means an empty world whose constituents are 
only unphysical objects.
Based on it, we speculate that
local fields emerge from unobservable ones
by unknown mechanisms.
Our question is how physical fields come from
the world with only a vacuum state as the physical state.

In this paper, we study a mechanism to release physical fields 
from unobservable ones, in the expectation 
that a useful hint on the origin of our world is provided.
We present a basic idea and a toy model
defined on a higher-dimensional space-time.
The model has fermionic symmetries, and
higher-dimensional fields form non-singlets under those
transformations and become unphysical.
We show that
some singlets can appear after the dimensional reduction
and become physical.

The outline of this paper is as follows.
In the next section, we give our basic idea.
In section 3, we present a toy model 
that physical modes come from unobservable ones
through a dimensional reduction.
In the last section, we give conclusions and discussions.

\section{Release of physical modes}
\label{Release of physical modes}

\subsection{Basic idea}
\label{Basic idea}

Our basic idea is summarized as follows.
We assume that the world just after the birth of space-time 
is effectively described by a theory with unphysical particles
on a higher dimensional space-time.

Let the fermionic conserved charges $Q_{\rm f}^A$ satisfy the algebraic relations,
\begin{eqnarray}
\{Q_{\rm f}^A, Q_{\rm f}^B\} = \sum_{i} f^{ABi} N^i~,~~
[N^i, Q_{\rm f}^A] = i \sum_{B} f^{iAB} Q_{\rm f}^B~,~~
[N^i, N^j] = i \sum_{k} f^{ijk} N^k~,
\label{Q-alg}
\end{eqnarray}
where $\{O_1, O_2\} \equiv O_1 O_2 + O_2 O_1$, $[O_1, O_2]=O_1 O_2 - O_2 O_1$
and $N^i$ are some bosonic conserved charges.
$f^{ABi}$, $f^{jAB}$ and $f^{ijk}$ are the structure constants 
that satisfy the relations,
\begin{eqnarray}
&~& \sum_i \left(f^{ABi} f^{iCD} + f^{BCi} f^{iAD} + f^{CAi} f^{iBD}\right) = 0~,~~
\label{J1-f}\\
&~& \sum_j f^{ABj} f^{jik} + \sum_C \left(f^{iBC} f^{CAk} + f^{iAC} f^{CBk}\right) = 0~,~~
\label{J2-f}\\
&~& \sum_k f^{ijk} f^{kAC} - \sum_B \left(f^{jAB} f^{iBC} + f^{iAB} f^{jBC}\right)= 0~,~~
\label{J3-f}\\
&~& \sum_l \left(f^{ijl} f^{lkm} + f^{jkl} f^{lim} + f^{kil} f^{ljm}\right) = 0~,
\label{J4-f}
\end{eqnarray}
from the Jacobi identities,
\begin{eqnarray}
&~& [\{Q_{\rm f}^A, Q_{\rm f}^B\}, Q_{\rm f}^C] + [\{Q_{\rm f}^B, Q_{\rm f}^C\}, Q_{\rm f}^A]
+  [\{Q_{\rm f}^C, Q_{\rm f}^A\}, Q_{\rm f}^B] = 0~,~~
\label{J1}\\
&~& [\{Q_{\rm f}^A, Q_{\rm f}^B\}, N^i] - \{[Q_{\rm f}^B, N^i], Q_{\rm f}^A\}
+ \{[N^i, Q_{\rm f}^A], Q_{\rm f}^B\} = 0~,~~
\label{J2}\\
&~& [[N^i, N^j], Q_{\rm f}^A] +[[N^j, Q_{\rm f}^A], N^i]
+ [[Q_{\rm f}^A, N^i], N^j] = 0~,~~
\label{J3}\\
&~& [[N^i, N^j], N^k] + [[N^j, N^k], N^i] + [[N^k, N^i], N^j] = 0~.
\label{J4}
\end{eqnarray}

The system is supposed to be schematically expressed by the Lagrangian density
$\mathcal{L}_{(4+n){\rm D}}$ or the Hamiltonian density $\mathcal{H}_{(4+n){\rm D}}$ such that
\begin{eqnarray}
&~& \mathcal{L}_{(4+n){\rm D}} 
= \bm{\delta}_{\rm f}^{(1)} \mathcal{R}_{(4+n){\rm D}}^{(1)}
= \bm{\delta}_{\rm f}^{(2)} \mathcal{R}_{(4+n){\rm D}}^{(2)}
= \cdots
= \bm{\delta}_{\rm f}^{(s)} \mathcal{R}_{(4+n){\rm D}}^{(s)}~,~~
\label{L-(4+n)D}\\
&~& \mathcal{H}_{(4+n){\rm D}} 
= i \{Q_{\rm f}^{(1)}, \tilde{\mathcal{R}}_{(4+n){\rm D}}^{(1)}\}
= i \{Q_{\rm f}^{(2)}, \tilde{\mathcal{R}}_{(4+n){\rm D}}^{(2)}\}
= \cdots
= i \{Q_{\rm f}^{(s)}, \tilde{\mathcal{R}}_{(4+n){\rm D}}^{(s)}\}~,
\label{H-(4+n)D}
\end{eqnarray}
where $\bm{\delta}_{\rm f}^{(r)}$ $(r=1, 2, \cdots, s)$ are defined by
$\zeta^{(r)} \bm{\delta}_{\rm f}^{(r)} \mathcal{O}_{(4+n){\rm D}} 
= i[\zeta^{(r)} Q_{\rm f}^{(r)}, \mathcal{O}_{(4+n){\rm D}}]$
with Grassmann parameters $\zeta^{(r)}$,
and $Q_{\rm f}^{(r)}$ are linear combinations of $Q_{\rm f}^A$, i. e.,
$Q_{\rm f}^{(r)} \equiv \sum_A a_A^{(r)} Q_{\rm f}^A$ 
with some constants $a_A^{(r)}$.
$\mathcal{O}_{(4+n){\rm D}} = \mathcal{O}_{(4+n){\rm D}}(x, y)$ is 
an operator on the higher-dimensional space-time,
where $x$ and $y$ stand for
the coordinates of a 4-dimensional (4D) space-time 
and an extra space, respectively.

Physical states denoted by $|{\rm phys}\rangle$ can be selected
by imposing the following conditions on states,
\begin{eqnarray}
Q_{\rm f}^{(r)} |{\rm phys}\rangle = 0~~~~~(r=1, 2, \cdots, s)~.
\label{Phys}
\end{eqnarray}
The conditions (\ref{Phys}) are interpreted 
as counterparts of the Kugo-Ojima subsidiary condition
in the BRST quantization~\cite{K&O1,K&O2,BRST}.

The system does not evolve because of the relation
$\langle {\rm phys} |\mathcal{H}_{(4+n){\rm D}}| {\rm phys} \rangle = 0$
derived from (\ref{H-(4+n)D}) and (\ref{Phys}).
Every field belongs to a member of non-singlets
under the fermionic charges $Q_{\rm f}^{(r)}$, i.e.,
every particle pairs with its ghost partner that is related to by
$Q_{\rm f}^{(r)}$, and it is unphysical.
Then, only the vacuum $| 0 \rangle$ survives as the physical state.

For the emergence of a physical mode, it is necessary 
to disappear its ghost partner.
Based on the orbifold breaking mechanism 
that some modes are eliminated by orbifolding the extra space~\cite{Orb},
we assume that the structure of space-time changes,
the configuration of fields are altered,
and then the boundary conditions of fields on the extra space
are determined dynamically.
After the dimensional reduction to the 4D space-time, 
the system is schematically expressed 
by the Lagrangian density $\mathcal{L}_{\rm 4D}$
or the Hamiltonian density $\mathcal{H}_{\rm 4D}$
such that, in $s'$ different way,
\begin{eqnarray}
&~& \mathcal{L}_{(4+n){\rm D}} \to 
\mathcal{L}_{\rm 4D} =  \mathcal{L} 
+ \bm{\delta}_{\rm f (4D)}^{(r')} \mathcal{R}_{\rm 4D}^{(r')}~,
\label{L-4D}\\
&~& \mathcal{H}_{(4+n){\rm D}} \to 
\mathcal{H}_{\rm 4D} =  \mathcal{H} 
+ i\{Q_{\rm f (4D)}^{(r')}, \tilde{\mathcal{R}}_{\rm 4D}^{(r')}\}~,
\label{H-4D}
\end{eqnarray}
where $\bm{\delta}_{\rm f(4D)}^{(r')}$ $(r'=1, 2, \cdots, s')$ are defined by
$\zeta^{(r')} \bm{\delta}_{\rm f(4D)}^{(r')} \mathcal{O}_{\rm 4D} 
= i[\zeta^{(r')} Q_{\rm f(4D)}^{(r')}, \mathcal{O}_{\rm 4D}]$
with Grassmann parameters $\zeta^{(r')}$ and
an operator $\mathcal{O}_{\rm 4D} = \mathcal{O}_{\rm 4D}(x)$
defined on the 4D space-time, 
and $Q_{\rm f(4D)}^{(r')}$ are fermionic charges.

Physical states can be selected by imposing the following conditions on states,
\begin{eqnarray}
{Q}_{\rm f(4D)}^{(r')} |{\rm phys}\rangle = 0~~~~~(r'=1, 2, \cdots, s')~.
\label{Phys2}
\end{eqnarray}

Unless $\mathcal{H}$ is written by an exact form for ${Q}_{\rm f(4D)}^{(r')}$,
$\mathcal{H}$ can contain 4D ${Q}_{\rm f(4D)}^{(r')}$ singlet fields,
i.e., ghost partnerless particles.
Then, physical states including them appear in the system.

\section{An example}
\label{An example}

\subsection{5-dimensional toy model}
\label{5D toy model}

We give a toy model defined on a 5-dimensional (5D) flat space-time.
Let us begin with the Lagrangian density,
\begin{eqnarray}
&~& \mathcal{L}_{\rm 5D} =\partial_M \varphi^{\dagger} \partial^{M} \varphi 
- m^2 \varphi^{\dagger} \varphi
+ \partial_M c_{\varphi}^{\dagger} \partial^{M} c_{\varphi}
- m^2 c_{\varphi}^{\dagger} c_{\varphi}
\nonumber \\
&~& ~~~~~~~~~ = \sum_{a=1}^{N} \left(\partial_M \varphi^{a \dagger} \partial^{M} \varphi^a 
- m^2 \varphi^{a \dagger} \varphi^a
+ \partial_M c_{\varphi}^{a \dagger} \partial^{M} c_{\varphi}^a
- m^2 c_{\varphi}^{a \dagger} c_{\varphi}^a\right)~,
\label{L-5D}
\end{eqnarray}
where $M=0,1,2,3,5$,
$\varphi^a=\varphi^a(x,y)$ are ordinary 5D complex scalar fields 
yielding the commutation relations,
and $c_{\varphi}^a=c_{\varphi}^a(x,y)$ are 5D complex scalar fields
yielding the anti-commutation relations.
Here $x^{\mu}$ $(\mu=0,1,2,3)$ and $x^5$ are denoted by $x$ and $y$, respectively.
Both $\varphi^a$ and $c^a_{\varphi}$ form $N$-plets of $U(N)$.

Based on the formulation with the property that
{\it the hermitian conjugate of canonical momentum
for a variable is just the canonical momentum for the
hermitian conjugate of the variable}~\cite{YK2},
we define the conjugate momenta of $\varphi$, $\varphi^{\dagger}$, 
$c_{\varphi}$ and $c_{\varphi}^{\dagger}$ as
\begin{eqnarray}
&~& \pi \equiv 
\left(\frac{\partial \mathcal{L}_{\rm 5D}}{\partial \dot{\varphi}}\right)_{\rm R}
=  \dot{\varphi}^{\dagger}~,~~
\pi^{\dagger} \equiv 
\left(\frac{\partial \mathcal{L}_{\rm 5D}}{\partial \dot{\varphi}^{\dagger}}\right)_{\rm L}
= \dot{\varphi}~,
\label{pi}\\
&~& \pi_{c_{\varphi}} \equiv 
\left(\frac{\partial \mathcal{L}_{\rm 5D}}{\partial \dot{c}_{\varphi}}\right)_{\rm R}
=  \dot{c}_{\varphi}^{\dagger}~,~~
\pi_{c_{\varphi}}^{\dagger} 
\equiv 
\left(\frac{\partial \mathcal{L}_{\rm 5D}}{\partial \dot{c}_{\varphi}^{\dagger}}\right)_{\rm L}
= \dot{c}_{\varphi}~,
\label{pi-c}
\end{eqnarray}
where R and L stand for the right-differentiation and the left-differentiation, respectively.

By solving the Klein-Gordon equations 
$\left(\raisebox{-0.6mm}{\LBox} + m^2\right)\varphi =0$
and $\left(\raisebox{-0.6mm}{\LBox} + m^2\right)c_{\varphi} =0$, 
we obtain the solutions,
\begin{eqnarray}
&~& \varphi(x) = \int \frac{d^4k}{\sqrt{(2\pi)^4 2k_0}}
\left(a(\bm{k}) e^{-i k x} + b^{\dagger}(\bm{k}) e^{i k x}\right)~,
\label{varphi-sol}\\
&~& \varphi^{\dagger}(x) = \int \frac{d^4k}{\sqrt{(2\pi)^4 2k_0}}
\left(a^{\dagger}(\bm{k}) e^{i k x} + b (\bm{k}) e^{-i k x}\right)~,
\label{varphi-dagger-sol}\\
&~& \pi(x) = i \int d^4k \sqrt{\frac{k_0}{2 (2\pi)^4}}
\left(a^{\dagger}(\bm{k}) e^{i k x} - b (\bm{k}) e^{-i k x}\right)~,
\label{pi-sol}\\
&~& \pi^{\dagger}(x) = - i \int d^4k \sqrt{\frac{k_0}{2 (2\pi)^4}}
\left(a(\bm{k}) e^{-i k x} - b^{\dagger} (\bm{k}) e^{i k x}\right)~,
\label{pi-dagger-sol}\\
&~& c_{\varphi}(x) = \int \frac{d^4k}{\sqrt{(2\pi)^4 2k_0}}
\left(c(\bm{k}) e^{-i k x} + d^{\dagger}(\bm{k}) e^{i k x}\right)~,
\label{c-sol}\\
&~& c_{\varphi}^{\dagger}(x) = \int \frac{d^4k}{\sqrt{(2\pi)^4 2k_0}}
\left(c^{\dagger}(\bm{k}) e^{i k x} + d (\bm{k}) e^{-i k x}\right)~,
\label{c-dagger-sol}\\
&~& \pi_{c_{\varphi}}(x) = i \int d^4k \sqrt{\frac{k_0}{2 (2\pi)^4}}
\left(c^{\dagger}(\bm{k}) e^{i k x} - d (\bm{k}) e^{-i k x}\right)~,
\label{pi-c-sol}\\
&~& \pi_{c_{\varphi}}^{\dagger}(x) = - i \int d^4k \sqrt{\frac{k_0}{2 (2\pi)^4}}
\left(c(\bm{k}) e^{-i k x} - d^{\dagger} (\bm{k}) e^{i k x}\right)~,
\label{pi-c-dagger-sol}
\end{eqnarray}
where $k_0 = \sqrt{\bm{k}^2 + m^2}$ and $kx = k^M x_M$.

The system is quantized by regarding variables as operators
and imposing the following commutation or anti-commutation relations 
on the canonical pairs,
\begin{eqnarray}
\hspace{-1cm}&~& [\varphi^a(\bm{x}, t), \pi^{a'}(\bm{y}, t)] 
= i \delta^{aa'} \delta^4(\bm{x}-\bm{y})~,~~ 
[\varphi^{a \dagger}(\bm{x}, t), \pi^{a' \dagger}(\bm{y}, t)] 
= i \delta^{aa'} \delta^4(\bm{x}-\bm{y})~,
\label{CCR-varphi}\\
\hspace{-1cm}&~& \{c_{\varphi}^a(\bm{x}, t), \pi_{c_{\varphi}}^{a'}(\bm{y}, t)\} 
= i \delta^{aa'} \delta^4(\bm{x}-\bm{y})~,~~ 
\{c_{\varphi}^{a \dagger}(\bm{x}, t), \pi_{c_{\varphi}}^{a' \dagger}(\bm{y}, t)\} 
= -i \delta^{aa'} \delta^4(\bm{x}-\bm{y})~,
\label{CCR-c}
\end{eqnarray}
and others are zero.
Or equivalently, for operators 
$a(\bm{k})$, $b^{\dagger}(\bm{k})$, $a^{\dagger}(\bm{k})$, $b(\bm{k})$,
$c(\bm{k})$, $d^{\dagger}(\bm{k})$, $c^{\dagger}(\bm{k})$ and $d(\bm{k})$,
the following relations are imposed on,
\begin{eqnarray}
&~& [a^a(\bm{k}), a^{a'\dagger}(\bm{l})] 
= \delta^{aa'} \delta^4(\bm{k}-\bm{l})~,~~ 
[b^a(\bm{k}), b^{a'\dagger}(\bm{l})] 
= \delta^{aa'} \delta^4(\bm{k}-\bm{l})~,~~ 
\label{CCR-ab-varphi}\\
&~& \{c^a(\bm{k}), c^{a'\dagger}(\bm{l})\} 
= \delta^{aa'} \delta^4(\bm{k}-\bm{l})~,~~
\{d^a(\bm{k}), d^{a'\dagger}(\bm{l})\} 
= - \delta^{aa'} \delta^4(\bm{k}-\bm{l})~,~~ 
\label{CCR-cd-c}
\end{eqnarray}
and others are zero.

The state vectors in the Fock space are constructed by operating
creation operators $a^{a\dagger}(\bm{k})$, $b^{a\dagger}(\bm{k})$,
$c^{a\dagger}(\bm{k})$ and $d^{a\dagger}(\bm{k})$ 
from the vacuum state $| 0 \rangle$
that satisfy $a^a(\bm{k}) | 0 \rangle = 0$, $b^a(\bm{k}) | 0 \rangle = 0$, 
$c^a(\bm{k}) | 0 \rangle = 0$ and $d^a(\bm{k}) | 0 \rangle = 0$.
Note that the system contains negative norm states
as seen from the relation $\{d^a(\bm{k}), d^{a'\dagger}(\bm{l})\} 
= - \delta^{aa'} \delta^4(\bm{k}-\bm{l})$.
For instance, from the relation,
\begin{eqnarray}
\hspace{-1cm}&~& 0 < \int d^4k \left|f^a(\bm{k})\right|^2 
= - \int d^4k \int d^4l f^{a*}(\bm{k}) f^a(\bm{l}) 
\langle 0 |\{d^a(\bm{k}), d^{a\dagger}(\bm{l})\}| 0 \rangle
\nonumber \\
\hspace{-1cm}&~& ~~~~~~~~~ = - \int d^4k \int d^4l f^{a*}(\bm{k}) f^a(\bm{l}) 
\langle 0 |d^a(\bm{k})d^{a\dagger}(\bm{l})| 0 \rangle
= - \left|\int d^4k f^a(\bm{k})d^{a\dagger}(\bm{k})| 0 \rangle\right|^2,
\label{f-square}
\end{eqnarray} 
we see that the state 
$\int d^4 k f^a(\bm{k})d^{a\dagger}(\bm{k})| 0 \rangle$ has a negative norm.
Here, $f^a(\bm{k})$ are some square integrable functions.

The $\mathcal{L}_{\rm 5D}$ is invariant 
under the $U(N)(\supset SU(N) \times U(1))$ transformation,
\begin{eqnarray}
\hspace{-1cm}&~& \delta^{\alpha} \varphi = i \epsilon^{\alpha} T^{\alpha} \varphi~,~~
\delta^{\alpha} \varphi^{\dagger} = - i \epsilon^{\alpha} \varphi^{\dagger}T^{\alpha}~,~~
\delta^{\alpha} c_{\varphi}  = i \epsilon^{\alpha} T^{\alpha} c_{\varphi}~,~~
\delta^{\alpha} c_{\varphi}^{\dagger} = - i \epsilon^{\alpha} c_{\varphi}^{\dagger} T^{\alpha}~,
\label{delta-SU(N)}\\
\hspace{-1cm}&~& \delta \varphi = i \epsilon \varphi~,~~
\delta \varphi^{\dagger} = - i \epsilon \varphi^{\dagger}~,~~
\delta c_{\varphi} = i \epsilon c_{\varphi}~,~~
\delta c_{\varphi}^{\dagger} = - i \epsilon c_{\varphi}^{\dagger}
\label{delta-U(1)}
\end{eqnarray}
and the fermionic transformations,
\begin{eqnarray}
&~& \delta_{\rm F}^{\alpha} \varphi 
= - \zeta^{\alpha} T^{\alpha} c_{\varphi}~,~~
\delta_{\rm F}^{\alpha} \varphi^{\dagger} = 0~,~~ 
\delta_{\rm F}^{\alpha} c_{\varphi} = 0~,~~
\delta_{\rm F}^{\alpha} c_{\varphi}^{\dagger} 
= \zeta^{\alpha} \varphi^{\dagger} T^{\alpha}~,
\label{delta-F-N}\\
&~& \delta_{\rm F} \varphi = - \zeta c_{\varphi}~,~~
\delta_{\rm F} \varphi^{\dagger} = 0~,~~ 
\delta_{\rm F} c_{\varphi} = 0~,~~
\delta_{\rm F} c_{\varphi}^{\dagger} = \zeta \varphi^{\dagger}~,
\label{delta-F}\\
&~& \delta_{\rm F}^{\alpha \dagger} \varphi = 0~,~~
\delta_{\rm F}^{\alpha \dagger} \varphi^{\dagger} 
= \zeta^{\alpha \dagger} c_{\varphi}^{\dagger} T^{\alpha}~,~~
\delta_{\rm F}^{\alpha \dagger} c_{\varphi} 
= \zeta^{\alpha \dagger} T^{\alpha} \varphi~,~~
\delta_{\rm F}^{\alpha \dagger} c_{\varphi}^{\dagger} = 0~,
\label{delta-Fdagger-N}\\
&~& \delta_{\rm F}^{\dagger} \varphi = 0~,~~
\delta_{\rm F}^{\dagger} \varphi^{\dagger} 
= \zeta^{\dagger} c_{\varphi}^{\dagger}~,~~
\delta_{\rm F}^{\dagger} c_{\varphi} = \zeta^{\dagger} \varphi~,~~
\delta_{\rm F}^{\dagger} c_{\varphi}^{\dagger} = 0~,
\label{delta-Fdagger}
\end{eqnarray}
where $\epsilon^{\alpha}$ $(\alpha = 1, 2, \cdots, N^2-1)$ 
and $\epsilon$ are infinitesimal real parameters, and
$\zeta^{\alpha}$ and $\zeta$ are Grassmann parameters.

The above transformations are generated by the conserved charges as follows,
\begin{eqnarray}
&~& \delta^{\alpha} \mathcal{O}_{\rm 5D} 
= i [\epsilon^{\alpha} N^{\alpha}, \mathcal{O}_{\rm 5D}]~,~~
\delta \mathcal{O}_{\rm 5D} = i [\epsilon N^0, \mathcal{O}_{\rm 5D}]~,~~
\label{U(N)}\\
&~& \delta_{\rm F}^{\alpha} \mathcal{O}_{\rm 5D} 
= i[\zeta^{\alpha} Q_{\rm F}^{\alpha}, \mathcal{O}_{\rm 5D}]~,~~
\delta_{\rm F} \mathcal{O}_{\rm 5D} 
= i[\zeta Q_{\rm F}, \mathcal{O}_{\rm 5D}]~,~~
\label{Fermionic}\\
&~& \delta_{\rm F}^{\alpha \dagger} \mathcal{O}_{\rm 5D} 
= i[Q_{\rm F}^{\alpha \dagger} \zeta^{\alpha \dagger},  \mathcal{O}_{\rm 5D}]~,~~
\delta_{\rm F}^{\dagger} \mathcal{O}_{\rm 5D} 
= i[Q_{\rm F}^{\dagger} \zeta^{\dagger},  \mathcal{O}_{\rm 5D}]~,
\label{Fermionic-dagger}
\end{eqnarray}
where $N^{\alpha}$ and $N^0$ are 
the $SU(N)$ and $U(1)$ conserved hermitian charges, 
and $Q_{\rm F}^{\alpha}$, $Q_{\rm F}$, 
$Q_{\rm F}^{\alpha \dagger}$ and $Q_{\rm F}^{\dagger}$
are the fermionic conserved charges.
Note that $\delta_{\rm F}$ and $\delta_{\rm F}^{\dagger}$
are not generated by hermitian operators, 
different from the generator of the BRST transformation
in systems with first class constraints~\cite{BRST} 
and that of the topological symmetry~\cite{W,Top}.

From $Q_{\rm F}^{\alpha}$, $Q_{\rm F}$, 
$Q_{\rm F}^{\alpha \dagger}$ and $Q_{\rm F}^{\dagger}$,
we can construct the fermionic conserved hermitian charges 
$Q_1^{\alpha}$, $Q_2^{\alpha}$, $Q_1$ and $Q_2$ such that
\begin{eqnarray}
Q_1^{\alpha} \equiv Q_{\rm F}^{\alpha} + Q_{\rm F}^{\alpha \dagger}~,~~
Q_2^{\alpha} \equiv i(Q_{\rm F}^{\alpha} - Q_{\rm F}^{\alpha \dagger})~,~~
Q_1 \equiv Q_{\rm F} + Q_{\rm F}^{\dagger}~,~~
Q_2 \equiv i(Q_{\rm F} - Q_{\rm F}^{\dagger})~.
\label{Q12}
\end{eqnarray}

The conserved charges satisfy the algebraic relations,
\begin{eqnarray}
\hspace{-1cm}&~& [N^{\alpha}, N^{\beta}] 
= i \sum_{\gamma=1}^{N^2-1} f^{\alpha\beta\gamma} N^{\gamma}~,~~
[N^{\alpha}, Q_1^{\beta}] =  i \sum_{\gamma=1}^{N^2-1} f^{\alpha\beta\gamma} Q_1^{\gamma}~,~~
[N^{\alpha}, Q_2^{\beta}] =  i \sum_{\gamma=1}^{N^2-1} f^{\alpha\beta\gamma} Q_2^{\gamma}~,~~
\nonumber \\
\hspace{-1cm}&~& 
\{Q_1^{\alpha}, Q_1^{\beta}\} = \sum_{\gamma=1}^{N^2-1} f^{\alpha\beta\gamma} N^{\gamma}~,~~ 
\{Q_2^{\alpha}, Q_2^{\beta}\} = \sum_{\gamma=1}^{N^2-1} f^{\alpha\beta\gamma} N^{\gamma}~,~~
\nonumber \\ 
\hspace{-1cm}&~& 
\{Q_1^{\alpha}, Q_2^{\beta}\} = \frac{1}{N} \delta^{\alpha\beta} N^0 + 
\sum_{\gamma=1}^{N^2-1} d^{\alpha\beta\gamma} N^{\gamma}~,~~
(Q_1)^2 = N^0~,~~ (Q_2)^2 = N^0~,~~ \{Q_1, Q_2\} = 0~,
\label{Q-alg-SU(N)}
\end{eqnarray}
where $f^{\alpha\beta\gamma}$ and $d^{\alpha\beta\gamma}$
are structure constants of the Lie algebra $su(N)$ that satisfy
the relations
$[T^{\alpha}, T^{\beta}] = i \sum_{\gamma=1}^{N^2-1} f^{\alpha\beta\gamma} T^{\gamma}$
and $\{T^{\alpha}, T^{\beta}\} =  \frac{1}{N} \delta^{\alpha\beta} I +
 i \sum_{\gamma=1}^{N^2-1} d^{\alpha\beta\gamma} T^{\gamma}$
($I$ is the $N \times N$ unit matrix),
and $N^0$ commutes to every charge.

The $\mathcal{L}_{\rm 5D}$ is rewritten as 
\begin{eqnarray}
\mathcal{L}_{\rm 5D} =  \bm{\delta}_{\rm F} \mathcal{R}_{\rm 5D}
=  \bm{\delta}_{\rm F}^{\dagger} \mathcal{R}_{\rm 5D}^{\dagger}
=  {\bm \delta}_{\rm F} {\bm \delta}_{\rm F}^{\dagger} \mathcal{L}_{\rm 5D}^{\varphi}
= - {\bm \delta}_{\rm F}^{\dagger} {\bm \delta}_{\rm F} \mathcal{L}_{\rm 5D}^{\varphi}~,
\label{L-5Dexact}
\end{eqnarray}
where $\bm{\delta}_{\rm F}$ and $\bm{\delta}_{\rm F}^{\dagger}$
are defined by $\delta_{\rm F} = \zeta \bm{\delta}_{\rm F}$
and $\delta_{\rm F}^{\dagger} = \zeta^{\dagger} \bm{\delta}_{\rm F}^{\dagger}$,
and
$\mathcal{R}_{\rm 5D}$, $\mathcal{R}_{\rm 5D}^{\dagger}$ 
and $\mathcal{L}_{\rm 5D}^{\varphi}$ are given by
\begin{eqnarray}
&~& \mathcal{R}_{\rm 5D} = \partial_M c_{\varphi}^{\dagger} \partial^{M} \varphi 
- m^2 c_{\varphi}^{\dagger} \varphi~,~~
\mathcal{R}_{\rm 5D}^{\dagger} = \partial_M \varphi^{\dagger} \partial^{M} c_{\varphi}
- m^2 \varphi^{\dagger} c_{\varphi}~,
\label{R-5D-dagger}\\
&~& \mathcal{L}_{\rm 5D}^{\varphi} = \partial_M \varphi^{\dagger} \partial^{M} \varphi
- m^2 \varphi^{\dagger} \varphi~,
\label{L-5D-varphi}
\end{eqnarray}
respectively.

The Hamiltonian density
$\mathcal{H}_{\rm 5D}$ is written in the $Q_{\rm F}$ and $Q_{\rm F}^{\dagger}$
exact forms such that 
\begin{eqnarray}
\mathcal{H}_{\rm 5D} =  i\{Q_{\rm F}, \tilde{\mathcal{R}}_{\rm 5D}\}
 =  -i \{Q_{\rm F}^{\dagger}, \tilde{\mathcal{R}}_{\rm 5D}^{\dagger}\}
= \{Q_{\rm F}, \{Q_{\rm F}^{\dagger}, \mathcal{H}_{\rm 5D}^{\varphi}\}\}
= - \{Q_{\rm F}^{\dagger}, \{Q_{\rm F}, \mathcal{H}_{\rm 5D}^{\varphi}\}\}~,
\label{H-5Dexact}
\end{eqnarray}
where $Q_{\rm F}$, $Q_{\rm F}^{\dagger}$,
$\tilde{\mathcal{R}}_{\rm 5D}$, $\tilde{\mathcal{R}}_{\rm 5D}^{\dagger}$ 
and $\mathcal{H}_{\rm 5D}^{\varphi}$ are given by
\begin{eqnarray}
\hspace{-1cm}&~& Q_{\rm F}  
= \int \left(-\pi c_{\varphi} + \varphi^{\dagger} \pi_{c_{\varphi}}^{\dagger}\right) d^4x~,~~
Q_{\rm F}^{\dagger} 
= \int \left(-c_{\varphi}^{\dagger} \pi^{\dagger} 
+ \pi_{c_{\varphi}} \varphi\right) d^4x~,
\label{QF-5D}\\
\hspace{-1cm}&~& \tilde{\mathcal{R}}_{\rm 5D} = \pi_{c_{\varphi}} \pi^{\dagger} 
+ \bm{\nabla} c_{\varphi}^{\dagger} \bm{\nabla} \varphi
+ m^2 c_{\varphi}^{\dagger} \varphi~,~~
\tilde{\mathcal{R}}_{\rm 5D}^{\dagger} = \pi \pi_{c_{\varphi}}^{\dagger} 
+ \bm{\nabla} \varphi^{\dagger} \bm{\nabla} c_{\varphi}
+ m^2 \varphi^{\dagger} c_{\varphi}~,~~
\label{tildeR-5D}\\
\hspace{-1cm}&~& \mathcal{H}_{\rm 5D}^{\varphi} =  \pi \pi^{\dagger} 
+ \bm{\nabla} \varphi^{\dagger} \bm{\nabla} \varphi
+ m^2 \varphi^{\dagger} \varphi~.
\label{H-5D-varphi}
\end{eqnarray}

To formulate our model in a consistent manner,
we use a feature that {\it a conserved charge can, in general,
be set to be zero as a subsidiary condition}.
We impose the following subsidiary conditions on states 
to select physical states,
\begin{eqnarray}
{Q}_{\rm F} |{\rm phys}\rangle = 0~~~ {\rm and} ~~~
{Q}_{\rm F}^{\dagger} |{\rm phys}\rangle = 0~,
\label{Phys-QF-5D}
\end{eqnarray}
Note that $Q_{\rm F}^{\dagger} |{\rm phys}\rangle = 0$ 
means $\langle {\rm phys}|Q_{\rm F}=0$,
and $N^{0} |{\rm phys}\rangle = 0$ is also imposed on
from the relation $\{{Q}_{\rm F}, {Q}_{\rm F}^{\dagger}\} = N^0$.
We find that all states, 
except for the ground state $|0 \rangle$, are
unphysical because they do not satisfy (\ref{Phys-QF-5D}).
This feature is understood as the quartet mechanism~\cite{K&O1,K&O2}.
The projection operator $P^{(n)}$ on the states with $n$ particles
is given by
\begin{eqnarray}
P^{(n)} = \frac{1}{n} \left(a^{\dagger} P^{(n-1)} a + b^{\dagger} P^{(n-1)} b + c^{\dagger} P^{(n-1)} c 
- d^{\dagger} P^{(n-1)} d \right)~~~~(n \ge 1)~,
\label{P(n)}
\end{eqnarray}
and is written by
\begin{eqnarray}
P^{(n)} =  i \left\{Q_{\rm F}, R^{(n)}\right\}~,
\label{P(n)2}
\end{eqnarray}
where $R^{(n)}$ is given by
\begin{eqnarray}
R^{(n)} = \frac{1}{n} \left(c^{\dagger} P^{(n-1)} a + b^{\dagger} P^{(n-1)} d\right)~~~~(n \ge 1)~.
\label{R(n)}
\end{eqnarray}
We find that any state with $n \ge 1$ is unphysical from the relation
$\langle {\rm phys}|P^{(n)}|{\rm phys}\rangle = 0$ for  $n \ge 1$.
Then, we understand that both $\varphi$ and $c_{\varphi}$ 
become unphysical,
and only $| 0 \rangle$ is physical.
This can be regarded as a field theoretical version 
of the Parisi-Sourlas mechanism~\cite{P&S}.

Using $Q_1 (\equiv Q_{\rm F} + Q_{\rm F}^{\dagger})$ and
$Q_2 (\equiv i(Q_{\rm F} - Q_{\rm F}^{\dagger}))$,
$\mathcal{H}_{\rm 5D}$ is rewritten as
\begin{eqnarray}
\mathcal{H}_{\rm 5D} =  i\{Q_1, \tilde{\mathcal{R}}_{1}\}
 = i \{Q_2, \tilde{\mathcal{R}}_2\}~,
\label{H-5Dexact-Q12}
\end{eqnarray}
where $\tilde{\mathcal{R}}_{1}$ and $\tilde{\mathcal{R}}_{2}$ are given by
\begin{eqnarray}
\tilde{\mathcal{R}}_{1} = \frac{1}{2}\left(\tilde{\mathcal{R}}_{\rm 5D}
+ \tilde{\mathcal{R}}_{\rm 5D}^{\dagger}\right)~,~~
\tilde{\mathcal{R}}_{2} = \frac{1}{2i}\left(\tilde{\mathcal{R}}_{\rm 5D}
- \tilde{\mathcal{R}}_{\rm 5D}^{\dagger}\right)~,
\label{tildeR12}
\end{eqnarray}
respectively.
We can select only the vacuum state as physical states 
by imposing the following subsidiary conditions on states, 
in place of (\ref{Phys-QF-5D}),
\begin{eqnarray}
{Q}_1 |{\rm phys}\rangle = 0~~~ {\rm and} ~~~
{Q}_2 |{\rm phys}\rangle = 0~.
\label{Phys-Q12-5D}
\end{eqnarray}

As seen from the relations 
$(Q_1)^2 = N^0$, $ (Q_2)^2 = N^0$ and $\{Q_1, Q_2\} = 0$
(or ${Q_{\rm F}}^2 = 0$, ${Q_{\rm F}^{\dagger}}^2 = 0$
and $\{Q_{\rm F}, Q_{\rm F}^{\dagger}\} = N^0$),
our fermionic charges $Q_1$ and $Q_2$ 
(or $Q_{\rm F}$ and $Q_{\rm F}^{\dagger}$)
are different from BRST and anti-BRST charges.
Though $Q_1$, $Q_2$ and $N^0$ form elements
of the $N=2$ (quantum mechanical) supersymmetry algebra~\cite{N=2},
our system does not possess the space-time supersymmetry,
because $N^0$ is not our Hamiltonian
$H_{\rm 5D} \equiv \int \mathcal{H}_{\rm 5D} d^4x$
but the $U(1)$ charge.
In this way, our fermionic symmetries are different from
the BRST symmetry and the space-time supersymmetry.

\subsection{Dimensional reduction}
\label{Dimensional reduction}

We show that singlets under fermionic transformations can appear 
through a dimensional reduction and become physical modes.

We assume that the structure of space-time changes
into $M^4 \times S^1/Z_2$.
Here, $M^4$ is the 4D Minkowski space-time 
and $S^1/Z_2$ is the 1-dimensional (1D) orbifold
obtained by dividing a circle $S^1$ with the $Z_2$ reflection of
5-th coordinate such as $y \to -y$ or $y \to 2 \pi R - y$
($R$ is the radius of $S^1$).

If we require that the Lagrangian density should be single-valued on $M^4 \times S^1/Z_2$,
the following boundary conditions are allowed,
\begin{eqnarray}
&~& \varphi(x,-y) = \eta^0 P_0 \varphi(x, y)~,~~ 
\varphi(x, 2\pi R-y) = \eta^1 P_1 \varphi(x, y)~,~~
\label{BC-varphi}\\
&~& c_{\varphi}(x,-y) = \eta_c^0 P_0 c_{\varphi}(x, y)~,~~ 
c_{\varphi}(x, 2\pi R-y) = \eta_c^1 P_1 c_{\varphi}(x, y)~,~~
\label{BC-c}
\end{eqnarray}
where $\eta^0$ and $\eta^1$ are intrinsic $Z_2$ parities of $\varphi$,
$\eta_c^0$ and $\eta_c^1$ are intrinsic $Z_2$ parities of $c_{\varphi}$,
and $P_0$ and $P_1$ are $N \times N$ matrices that satisfy
$(P_0)^2 = I$ and $(P_1)^2 = I$, respectively.

We assume that the boundary conditions are determined
by an unknown mechanism,
and take  $(\eta^0, \eta^1)=(1, 1)$, $(\eta_c^0, \eta_c^1)=(-1, -1)$, and 
\begin{eqnarray}
P_0 = {\rm diag}(\underbrace{1, \cdots, 1}_{N})~,~~
P_1 = {\rm diag}(\underbrace{1, \cdots, 1}_{k}, \underbrace{-1, \cdots, -1}_{N-k})~.
\label{P01}
\end{eqnarray}
Then, $\varphi$ and $c_{\varphi}$ are given by the Fourier expansions,
\begin{eqnarray}
&~& \varphi^{a_+}(x,y) = \frac{1}{\sqrt{\pi R}} \varphi_0^{a_+}(x) 
+ \sqrt{\frac{2}{\pi R}} \sum_{n=1}^{\infty} \varphi_n^{a_+}(x) \cos\left(\frac{n}{R} y\right)~,
\label{varphia+}\\
&~& \varphi^{a_-}(x,y) 
= \sqrt{\frac{2}{\pi R}} \sum_{n=1}^{\infty} \varphi_n^{a_-}(x) \cos\left(\frac{n-\frac{1}{2}}{R} y\right)~,
\label{varphia-}\\
&~& c_{\varphi}^{a_+}(x,y) =\sqrt{\frac{2}{\pi R}} \sum_{n=1}^{\infty} c_n^{a_+}(x) 
\sin\left(\frac{n}{R} y\right)~,
\label{ca+}\\
&~& c_{\varphi}^{a_-}(x,y) 
= \sqrt{\frac{2}{\pi R}} \sum_{n=1}^{\infty} c_n^{a_-}(x) \sin\left(\frac{n-\frac{1}{2}}{R} y\right)~,
\label{ca-}
\end{eqnarray}
where $\varphi_0^{a_+}(x)$, $\varphi_n^{a_{\pm}}(x)$ and $c_n^{a_{\pm}}(x)$ 
$(a_+ = 1, \cdots, k, a_- = k+1, \cdots, N, n=1, 2, \cdots)$ are 4D fields.
Note that the fermionic symmetries are broken down explicitly
by the above intrinsic $Z_2$ parity assignments.

After inserting the expansions (\ref{varphia+}) -- (\ref{ca-}) into (\ref{L-5D})
and integrating the 5-th coordinate, we obtain the 4D Lagrangian density,
\begin{eqnarray}
&~& \mathcal{L}_{\rm 4D} = 
\sum_{a_+=1}^k \left(\partial_{\mu} \varphi_0^{a_+ \dagger} \partial^{\mu} \varphi_0^{a_+}
- m^2  \varphi_0^{a_+ \dagger} \varphi_0^{a_+}\right)
\nonumber \\
&~& ~~~~~~~~~~~~~ 
+ \sum_{a_+=1}^k
\sum_{n=1}^{\infty} \left[\partial_{\mu} \varphi_n^{a_+ \dagger} \partial^{\mu} \varphi_n^{a_+}
- \left(m^2 + \left(\frac{n}{R}\right)^2\right) \varphi_n^{a_+ \dagger} \varphi_n^{a_+} \right.
\nonumber \\
&~& ~~~~~~~~~~~~~~~~~~~~~~~~~~~~~~~~
\left.
+ \partial_{\mu} c_n^{a_+ \dagger} \partial^{\mu} c_n^{a_+}
- \left(m^2 + \left(\frac{n}{R}\right)^2\right) c_n^{a_+ \dagger} c_n^{a_+}\right]
\nonumber \\
&~& ~~~~~~~~~~~~~ 
+ \sum_{a_-=k+1}^N
\sum_{n=1}^{\infty} \left[\partial_{\mu} \varphi_n^{a_- \dagger} \partial^{\mu} \varphi_n^{a_-}
- \left(m^2 + \left(\frac{n-\frac{1}{2}}{R}\right)^2\right) \varphi_n^{a_- \dagger} \varphi_n^{a_-}
\right.
\nonumber \\
&~& ~~~~~~~~~~~~~~~~~~~~~~~~~~~~~~~
\left.
+ \partial_{\mu} c_n^{a_- \dagger} \partial^{\mu} c_n^{a_-}
- \left(m^2 + \left(\frac{n-\frac{1}{2}}{R}\right)^2\right) c_n^{a_- \dagger} c_n^{a_-}\right]~.
\label{L-4D-toy}
\end{eqnarray}

The $\mathcal{L}_{\rm 4D}$ is invariant under 
the $SU(k) \times SU(N-k) \times U(1)$ transformation,
\begin{eqnarray}
\hspace{-1cm}&~& \delta^{\hat{\alpha}} \varphi_0 = i \epsilon^{\hat{\alpha}} T^{\hat{\alpha}} \varphi_0~,~~
\delta^{\hat{\alpha}} \varphi_0^{\dagger} 
= - i \epsilon^{\hat{\alpha}} \varphi_0^{\dagger}T^{\hat{\alpha}}~,~~
\delta \varphi_0 = i \epsilon \varphi_0~,~~
\delta \varphi_0^{\dagger} = - i \epsilon \varphi_0^{\dagger}~,
\label{delta-varphi0} \\
\hspace{-1cm}&~& \delta^{\hat{\alpha}} \varphi_n 
= i \epsilon^{\hat{\alpha}} T^{\hat{\alpha}} \varphi_n~,~~
\delta^{\hat{\alpha}} \varphi_n^{\dagger} 
= - i \epsilon^{\hat{\alpha}} \varphi_n^{\dagger}T^{\hat{\alpha}}~,~~
\delta^{\hat{\alpha}} c_n  = i \epsilon^{\hat{\alpha}} T^{\hat{\alpha}} c_n~,~~
\delta^{\hat{\alpha}} c_n^{\dagger} 
= - i \epsilon^{\hat{\alpha}} c_n^{\dagger} T^{\hat{\alpha}}~,
\label{delta-SU(k)-n}\\
\hspace{-1cm}&~& \delta \varphi_n = i \epsilon \varphi_n~,~~
\delta \varphi_n^{\dagger} = - i \epsilon \varphi_n^{\dagger}~,~~
\delta c_n = i \epsilon c_n~,~~
\delta c_n^{\dagger} = - i \epsilon c_n^{\dagger}~,
\label{delta-U(1)-n}
\end{eqnarray}
where $\epsilon^{\hat{\alpha}}$ $(\hat{\alpha} = 1, 2, \cdots, k^2 + (N-k)^2 -2)$ and $\epsilon$ 
are infinitesimal real parameters,
$T^{\hat{\alpha}}$ are the elements of Lie algebra 
concerning $SU(k) \times SU(N-k)$, and
$\varphi_0$, $\varphi_n$ and $c_n$ stand for
multiplets $\varphi_0^{a_+}$, $\varphi_n^a$ and $c_n^a$
$(a=1,2,\cdots, N)$, respectively.

The $\mathcal{L}_{\rm 4D}$ is also invariant under the fermionic transformations,
\begin{eqnarray}
&~& \delta_{\rm F}^{\hat{\alpha}} \varphi_n 
= - \zeta^{\hat{\alpha}} T^{\hat{\alpha}} c_n~,~~
\delta_{\rm F}^{\hat{\alpha}} \varphi_n^{\dagger} = 0~,~~ 
\delta_{\rm F}^{\hat{\alpha}} c_n = 0~,~~
\delta_{\rm F}^{\hat{\alpha}} c_n^{\dagger} 
= \zeta^{\hat{\alpha}} \varphi_n^{\dagger} T^{\hat{\alpha}}~,
\label{delta-F-alpha-n}\\
&~& \delta_{\rm F} \varphi_n = - \zeta c_n~,~~
\delta_{\rm F} \varphi_n^{\dagger} = 0~,~~ 
\delta_{\rm F} c_n = 0~,~~
\delta_{\rm F} c_n^{\dagger} = \zeta \varphi_n^{\dagger}~,
\label{delta-F-n}\\
&~& \delta_{\rm F}^{\hat{\alpha} \dagger} \varphi_n = 0~,~~
\delta_{\rm F}^{\hat{\alpha} \dagger} \varphi_n^{\dagger} 
= \zeta^{\hat{\alpha} \dagger} c_n^{\dagger} T^{\hat{\alpha}}~,~~
\delta_{\rm F}^{\hat{\alpha} \dagger} c_n 
= \zeta^{\hat{\alpha} \dagger} T^{\hat{\alpha}} \varphi_n~,~~
\delta_{\rm F}^{\hat{\alpha} \dagger} c_n^{\dagger} = 0~,
\label{delta-Fdagger-alpha-n}\\
&~& \delta_{\rm F}^{\dagger} \varphi_n = 0~,~~
\delta_{\rm F}^{\dagger} \varphi_n^{\dagger} = \zeta^{\dagger} c_n^{\dagger}~,~~
\delta_{\rm F}^{\dagger} c_n = \zeta^{\dagger} \varphi_n~,~~
\delta_{\rm F}^{\dagger} c_n^{\dagger} = 0~,
\label{delta-Fdagger-n}
\end{eqnarray}
where $\zeta^{\hat{\alpha}}$ and $\zeta$ are Grassmann parameters.

The $\mathcal{L}_{\rm 4D}$ is rewritten by
\begin{eqnarray}
&~& \mathcal{L}_{\rm 4D} =  
\sum_{a_+=1}^{k} \left(\partial_{\mu} \varphi_0^{a_+ \dagger} \partial^{\mu} \varphi_0^{a_+}
- m^2  \varphi_0^{a_+ \dagger} \varphi_0^{a_+}\right)
+ \mathcal{L}_{\rm KK}~,~~
\nonumber \\
&~& \mathcal{L}_{\rm KK} 
= \bm{\delta}_{\rm F} \mathcal{R}_{\rm 4D}
= \bm{\delta}_{\rm F}^{\dagger} \mathcal{R}_{\rm 4D}^{\dagger}
= \bm{\delta}_{\rm F}  
\bm{\delta}_{\rm F}^{\dagger} \mathcal{L}_{\rm KK}^{\varphi}
= - \bm{\delta}_{\rm F}^{\dagger} \bm{\delta}_{\rm F} 
\mathcal{L}_{\rm KK}^{\varphi}~,
\label{L-4D-toy-rewritten}
\end{eqnarray}
where $\bm{\delta}_{\rm F}$ and $\bm{\delta}_{\rm F}^{\dagger}$
are defined by $\delta_{\rm F} = \zeta \bm{\delta}_{\rm F}$
and $\delta_{\rm F}^{\dagger} = \zeta^{\dagger} \bm{\delta}_{\rm F}^{\dagger}$, and
$\mathcal{R}_{\rm 4D}$ and $\mathcal{L}_{\rm KK}^{\varphi}$ are given by
\begin{eqnarray}
&~& \mathcal{R}_{\rm 4D} = \sum_{a_+=1}^{k} \sum_{n=1}^{\infty} 
\left[\partial_{\mu} c_n^{a_+ \dagger} \partial^{\mu} \varphi_n^{a_+}
- \left(m^2 + \left(\frac{n}{R}\right)^2\right) c_n^{a_+ \dagger} \varphi_n^{a_+}\right]
\nonumber \\
&~& ~~~~~~~~~~~~ 
+ \sum_{a_-=k+1}^N \sum_{n=1}^{\infty} 
\left[\partial_{\mu} c_n^{a_- \dagger} \partial^{\mu} \varphi_n^{a_-}
- \left(m^2 + \left(\frac{n-\frac{1}{2}}{R}\right)^2\right) c_n^{a_- \dagger} \varphi_n^{a_-}\right]~.
\label{R-4D}\\
&~& \mathcal{L}_{\rm KK}^{\varphi} = \sum_{a_+=1}^{k} \sum_{n=1}^{\infty} 
\left[\partial_{\mu} \varphi_n^{a_+ \dagger} \partial^{\mu} \varphi_n^{a_+}
- \left(m^2 + \left(\frac{n}{R}\right)^2\right) \varphi_n^{a_+ \dagger} \varphi_n^{a_+}\right]
\nonumber \\
&~& ~~~~~~~~~~~~ 
+ \sum_{a_-=k+1}^N \sum_{n=1}^{\infty} 
\left[\partial_{\mu} \varphi_n^{a_- \dagger} \partial^{\mu} \varphi_n^{a_-}
- \left(m^2 + \left(\frac{n-\frac{1}{2}}{R}\right)^2\right) \varphi_n^{a_- \dagger} \varphi_n^{a_-}\right]~.
\label{L-4D-varphi}
\end{eqnarray}
In the similar way, the Hamiltonian density $\mathcal{H}_{\rm 4D}$ is written by
\begin{eqnarray}
&~& \mathcal{H}_{\rm 4D} = \sum_{a_+=1}^{k} \left(\pi_0^{a_+} \pi_0^{a_+ \dagger} 
+ \bm{\nabla} \varphi_0^{a_+ \dagger} \bm{\nabla} \varphi_0^{a_+}
+ m^2 \varphi_0^{a_+ \dagger} \varphi_0^{a_+}\right)
+ \mathcal{H}_{\rm KK}~,~~
\nonumber \\
&~& \mathcal{H}_{\rm KK} 
= i \{{Q}_{\rm F(4D)}, \tilde{\mathcal{R}}_{\rm 4D}\}
= - i \{{Q}_{\rm F(4D)}^{\dagger}, \tilde{\mathcal{R}}_{\rm 4D}^{\dagger}\}
= \{{Q}_{\rm F(4D)}, 
\{{Q}_{\rm F(4D)}^{\dagger}, \mathcal{H}_{\rm KK}^{\varphi}\}\}
\nonumber \\
&~& ~~~~~~~~= - \{{Q}_{\rm F(4D)}^{\dagger}, 
\{{Q}_{\rm F(4D)}, \mathcal{H}_{\rm KK}^{\varphi}\}\}~,
\label{H-4D-toy-rewritten}
\end{eqnarray}
where ${Q}_{\rm F(4D)}$, $\tilde{\mathcal{R}}_{\rm 4D}$ and $\mathcal{H}_{\rm KK}^{\varphi}$
are given by
\begin{eqnarray}
&~& {Q}_{\rm F(4D)} = \int \sum_{a=1}^N \sum_{n=1}^{\infty} 
\left[-\pi_{n}^{a} c_{n}^{a}  + \varphi_n^{a \dagger} \pi_{c n}^{a \dagger} 
\right] d^3x~,~~
\label{tildeQF}\\
&~& \tilde{\mathcal{R}}_{\rm 4D} = \sum_{a_+=1}^{k} \sum_{n=1}^{\infty} 
\left[\pi_{c n}^{a_+ \dagger} \pi_{n}^{a_+} + \bm{\nabla} c_n^{a_+ \dagger} \bm{\nabla} \varphi_n^{a_+}
+ \left(m^2 + \left(\frac{n}{R}\right)^2\right) c_n^{a_+ \dagger} \varphi_n^{a_+}\right]
\nonumber \\
&~& ~~~~~~~~~~~~
+ \sum_{a_- = k+1}^{N} \sum_{n=1}^{\infty} 
\left[\pi_{c n}^{a_- \dagger} \pi_{n}^{a_-} + \bm{\nabla} c_n^{a_- \dagger} \bm{\nabla} \varphi_n^{a_-}
+ \left(m^2 + \left(\frac{n-\frac{1}{2}}{R}\right)^2\right)  c_n^{a_- \dagger} \varphi_n^{a_-}\right]~,
\label{tildeR-4D}\\
&~& \mathcal{H}_{\rm KK}^{\varphi} = \sum_{a_+=1}^{k} \sum_{n=1}^{\infty} 
\left[\pi_{n}^{a_+ \dagger} \pi_{n}^{a_+} + \bm{\nabla} \varphi_n^{a_+ \dagger} \bm{\nabla} \varphi_n^{a_+}
+ \left(m^2 + \left(\frac{n}{R}\right)^2\right) \varphi_n^{a_+ \dagger} \varphi_n^{a_+}\right]
\nonumber \\
&~& ~~~~~~~~~~~~
+ \sum_{a_- = k+1}^{N} \sum_{n=1}^{\infty} 
\left[\pi_{n}^{a_- \dagger} \pi_{n}^{a_-} + \bm{\nabla} \varphi_n^{a_- \dagger} \bm{\nabla} \varphi_n^{a_-}
+ \left(m^2 + \left(\frac{n-\frac{1}{2}}{R}\right)^2\right) \varphi_n^{a_- \dagger} \varphi_n^{a_-}\right]~.
\label{H-4D-varphi}
\end{eqnarray}
Here, $\pi_{0}^{a_+}  = \dot{\varphi}_{0}^{a_+ \dagger}$,
$\pi_{0}^{a_+ \dagger}  = \dot{\varphi}_{0}^{a_+}$,
$\pi_{n}^{a}  = \dot{\varphi}_{n}^{a \dagger}$,
$\pi_{n}^{a \dagger}  = \dot{\varphi}_{n}^{a}$,
$\pi_{cn}^{a}  = \dot{c}_{n}^{a \dagger}$
and
$\pi_{cn}^{a \dagger}  = \dot{c}_{n}^{a}$.

We impose the following subsidiary conditions on states,
\begin{eqnarray}
{Q}_{\rm F(4D)} |{\rm phys}\rangle = 0~~~ {\rm and} ~~~
{Q}_{\rm F(4D)}^{\dagger} |{\rm phys}\rangle = 0~,
\label{Phys-QF-4D}
\end{eqnarray}
and select physical states.
From (\ref{delta-F-alpha-n}) -- (\ref{delta-Fdagger-n}),
Kaluza-Klein (KK) modes $\varphi_n^a(x)$ and $c_n^a(x)$ $(n=1, 2, \cdots)$ 
form non-singlets of fermionic symmetries and become unphysical.
In contrast, $\varphi_0^{a_+}(x)$ are singlets of fermionic symmetries 
and become physical fields.
In this way, it is shown that the modes $\varphi_0^{a_+}(x)$ 
release from the unphysical fields
$\varphi^a(x,y)$, after the dimensional reduction.

We point out that our fermionic symmetries generated by 
${Q}_{\rm F(4D)}$ and ${Q}_{\rm F(4D)}^{\dagger}$
are regarded as accidental ones appearing after the compactification,
because the original fermionic symmetries are broken down
explicitely by the boundary conditions
as seen from the fact that
$Q_{\rm F}$ and $Q_{\rm F}^{\dagger}$
defined by (\ref{QF-5D})
vanish using the Fourier expansions (\ref{varphia+}) -- (\ref{ca-}).

\section{Conclusions and discussions}
\label{Conclusions and discussions}

We have studied a mechanism to release physical fields 
from unphysical ones, 
based on the conjecture 
that our world comes into existence from nothing.
We have proposed the idea
that physical modes can appear through the dimensional reduction
from unobservable particles on a higher-dimensional space-time,
and presented the toy model with fermionic symmetries
that make 5D fields unphysical forming non-singlets of those symmetries,
and 4D singlet modes come from through the compactification
with the orbifold breaking mechanism.\footnote{
As an attempt different from the standard lore,
toy models of fermions have been presented with
the feature that a finite numbers of Kaluza-Klein modes survive
without inducing masses after the integration over 
extra coordinates by imposing the invariance 
under space-time reflections and a shift relating extra space
on the Lagrangian density~\cite{Erdem1,Erdem2}.
}

There are many subjects left behind to explore the origin of
our world.

First one is to explore how the topology change occurs
and how boundary conditions are determined.
It is interesting to investigate a dynamical determination of
the structure of space-time and the pattern of boundary conditions.

Second one is to construct a realistic theory 
including the standard model as a low-energy theory
and a fundamental theory at a higher-energy scale, 
based on our conjecture and idea.
Larger fermionic symmetries would be needed to formulate unphysical theories
including gauge bosons and gravitons.
It is challenging to construct an interacting model
containing our coexisting system as a subsystem,
after an example of the gauge fixing term 
and the Faddeev-Popov ghost term~\cite{F&P}
in gauge theories~\cite{K&O1,K&O2,BRST}
and non-gauge model with BRST scalar doublets~\cite{F2,F3}.

Last one is to answer the question whether
our scenario is verified or not experimentally.
The first step is to find an experimental signature
for the existence of $Q_{\rm F}$-doublets.
For a system that $Q_{\rm F}$-singlets
and $Q_{\rm F}$-doublets coexist and interact with 
in the exact fermionic invariant way,
the Lagrangian density is, in general, written schematically such that
$\mathcal{L}_{\rm Total} = \mathcal{L}_{\rm S} + \mathcal{L}_{\rm D} + \mathcal{L}_{\rm mix}
= \mathcal{L}_{\rm S} + \bm{\delta}_{\rm F} \bm{\delta}_{\rm F}^{\dagger} (\Delta \mathcal{L})$.
Here, $\mathcal{L}_{\rm S}$, $\mathcal{L}_{\rm D}$ 
and $\mathcal{L}_{\rm mix}$
are the Lagrangian density for $Q_{\rm F}$-singlets, $Q_{\rm F}$-doublets
and interactions between $Q_{\rm F}$-singlets and $Q_{\rm F}$-doublets.
Under the subsidiary conditions 
$Q_{\rm F} |{\rm phys}\rangle =0$ and
$Q_{\rm F}^{\dagger} |{\rm phys}\rangle =0$
on states, all $Q_{\rm F}$-doublets become unphysical
and would not give any dynamical effects on $Q_{\rm F}$-singlets.
The system seems to be same as that described by 
$\mathcal{L}_{\rm S}$ alone.
From this, we suppose that it is not possible to show 
the existence of $Q_{\rm F}$-doublets.
However, in a very special case, an indirect proof would be possible
through fingerprints left by symmetries in a fundamental theory. 
The fingerprints are specific relations among parameters
such as a coupling unification,
reflecting on underlying symmetries~\cite{YK1}.
This subject will be reexamined in the separate publication~\cite{YK3}.

These studies would shed new light on the origin of our space-time
and the standard model, and provide us a hint on
the structure of ultimate theory.

\section*{Acknowledgments}
The author thanks Prof. T. Kugo for valuable discussions
and useful comments.
This work was supported in part by scientific grants from the Ministry of Education, Culture,
Sports, Science and Technology under Grant No.~22540272.

\end{document}